\documentstyle[aps,epsfig]{revtex}

\begin{document}


\title{Optical absorption in Fibonacci lattices at finite temperature}

\wideabs{
\author{A.\ Rodr\'{\i}guez}
\address{Departamento de Matem\'{a}tica Aplicada y Estad\'{\i}stica and
Grupo Interdisciplinar de Sistemas Complicados, Universidad
Polit\'{e}cnica, E-28040 Madrid, Spain}

\author{F.\ Dom\'{\i}nguez-Adame}
\address{Departamento de F\'{\i}sica de Materiales and Grupo
Interdisciplinar de Sistemas Complicados, Universidad Complutense,
E-28040 Madrid, Spain}

\maketitle

\begin{abstract}

We consider the dynamics of Frenkel excitons on quasiperiodic lattices,
focusing our attention on the Fibonacci case as a typical example.  We
evaluate the absorption spectrum by solving numerically the equation of
motion of the Frenkel-exciton problem on the lattice.  Besides the main
absorption line, satellite lines appear in the high-energy side of the
spectra, which we have related to the underlying quasiperiodic order.
The influence of lattice vibrations on the absorption line-shape is also
considered.  We find that the characteristic features of the absorption
spectra should be observable even at room temperature.  Consequently, we
propose that excitons act as a probe of the topology of the lattice even
when thermal vibrations reduce their quantum coherence.

\end{abstract}

}


\section{Introduction}

Low-dimensional quasiperiodic systems have been the subject of intensive
theoretical studies because of their unusual properties, such as fractal
energy spectra and self-similar wave functions.  Several electronic
properties of these systems can be inferred from optical measurements:
In particular, optical absorption techniques are suitable for their
characterization because cuasiperiodic order causes the occurrence of
well-defined lines which do not arise in periodic or random systems
\cite{PLA}.  In previous works~\cite{PLA,Macia94,Adame95,Adame96} we
have focused our attention on Frenkel excitons in Fibonacci lattices as
typical examples of low-dimensional quasiperiodic systems.  By solving
numerically the equation of motion of Frenkel excitons on the lattice,
in which on-site energies take on two values following the Fibonacci
sequence, we found that the characteristic satellites observed in the
high-energy side of the absorption spectra correspond to well-defined
peaks of the Fourier transform of the lattice.  This result is important
since it enables us to find a simple relationship between the underlying
long range order and optical measurements.  Due to the long-range nature
of Fibonacci order, it is clear that quantum coherence of excitons over
large distances is required to observe these characteristic features.
In most systems, however, there are inelastic scattering mechanisms
which could result in a reduction of the coherence length.  Among them,
the effect of lattice vibrations on the optical absorption line-shape
has long been a subject of special interest~\cite{Schreiber82}.

In this paper we investigate optical absorption due to Frenkel excitons
in Fibonacci lattices at finite temperature.  To this end, we consider
the Frenkel Hamiltonian and include the exciton-phonon interaction.  We
follow closely the approach previously introduced by Schreiber and
Toyozawa~\cite{Schreiber82} in periodic lattices to take into account
exciton-phonon interaction.  We use a general treatment to study the
dynamics of Frenkel excitons in these lattices, solve the microscopic
equation of motion proposed by Huber and Ching~\cite{Huber89} and find
the optical absorption spectrum.  The optical line-shape at different
temperatures is obtained.  The main aim of the paper is to demonstrate
that the satellite peaks appearing in the high energy portion of the
spectrum \cite{PLA}, which characterize the long-range order of the
lattice, are still present and observable at finite temperature.  If so,
we will have shown that excitons act as a probe of the lattice structure
even at finite temperature.

\section{Physical model at zero temperature}

We consider a system of $N$ optically active centers occupying positions on 
a linear regular lattice. The Frenkel Hamiltonian describing this system 
can be written as follows (we use units such that $\hbar=1$)
\begin{equation}
{\cal H}= \sum_{n}\> V_{n} a_{n}^{\dag}a_{n} -
J\sum_{n}\>(a_{n}^{\dag}a_{n+1}+a_{n+1}^{\dag}a_{n}).
\label{Hamiltonian}
\end{equation}
Here $a_{n}^{\dag}$ ($a_{n}$) creates (annihilates) an electronic
excitation of energy $V_n$ at site $n$, and $J$ is the hopping integral.
Any arbitrary Fibonacci system presents two kind of building blocks.  In
our case, we choose those blocks as individual centers A and B, with
$V_{n}=V_A$ or $V_{n}=V_B$~\cite{PLA}.  The Fibonacci sequence $S_n$ is
generated by appending the $n-2$ sequence to the $n-1$ one, i.e.,
$S_n=\{S_{n-1}S_{n-2}\}$.  This construction algorithm requires the
initial conditions $S_0=B$ and $S_1=A$.  The $n$ sequence has $N=F_n$
elements, where $F_n$ denotes the $n\,$th Fibonacci number satisfying
$F_n=F_{n-1} + F_{n-2}$ with $F_0=F_1=1$.  As $n$ increases the ratio
$F_{n-1}/F_n$ converges to $\tau = (\sqrt{5}-1)/2 = 0.618\ldots$, an
irrational number which is known as the inverse golden mean.  Therefore,
lattice centers are arranged according to the Fibonacci sequence
$ABAABABA\ldots$, where the fraction of $B$-centers is $c\sim 1-\tau$.

The line-shape $I(E)$ of an optical-absorption process in which a single
exciton is created in a lattice with $N$ centers can be obtained
as~\cite{Huber89}
\begin{equation}
I(E)=-\,{2\over \pi N} \int_0^\infty\> dt\, e^{-\alpha t} \sin (Et)
\mbox{Im}\left( \sum_{n} G_{n} (t) \right),
\label{spectrum}
\end{equation}
where the factor $\exp(-\alpha t)$ takes into account the broadening due
to the instrumental resolution function of half-width $\alpha$ and the
correlation functions $G_{n}(t)$ obey the equation of motion
\begin{equation}
i{d\over dt} G_{n}(t) = \sum_{l}\> H_{nl} G_{l}(t),
\label{motion}
\end{equation}
with the initial condition $G_{n}(0)=1$.  The diagonal elements of the
tridiagonal matrix $H_{nl}$ are $V_{n}$ whereas off-diagonal elements
are simply given by $-J$.

\section{Exciton-phonon problem}

In the previous section we have summarized the line-shape problem at
zero temperature.  Now we deal with the exciton-phonon coupling and its
influence on the optical absorption spectra.  Under lattice vibrations
the on-site energy will be subject to spatial and temporal
fluctuations~\cite{Schreiber82} around their zero-temperature values,
$V_A$ and $V_B$.  By means of the Franck-Condon approximation and
assuming that vibrational modes at different sites are independent,
Schreiber and Toyozawa~\cite{Schreiber82} cast the problem into an
equivalent one with random onsite correlations
\begin{equation}
\langle V_{n}V_{n^{\prime}} \rangle = D^{2}\delta_{nn^{\prime}}.
\label{delta}
\end{equation}
The approximations involved in writting (\ref{delta}) are correct in
several cases of interest, such as optical modes in inorganic crystals.
Therefore, we will consider in what follows the finite temperature
problem by using random on-site energies with Gaussian distribution
\begin{equation}
P(V_{n})={1\over 2\pi D^{2}} \exp \left( -\,{(V_{n}-\langle V
\rangle_{n})^{2}\over 2D^{2}} \right),
\label{distribution}
\end{equation}
$\langle V \rangle_{n}$ being either $V_A$ or $V_B$ according to the
Fibonacci sequence. Here $D^{2}=4gJk_BT$ in one-dimension, where $g$ is
the exciton-phonon coupling and $k_B$ is the Boltzmann's constant.

\section{Numerical results and discussions}

We have solved numerically the equation of motion (\ref{motion}) with
random on-site energies distributed according to (\ref{distribution}).
As typical parameters we have chosen $V_A=4.0\,$eV, $V_B=4.8\,$eV and
$J=0.5\,$eV so that the exciton bandwidth is $W=4J=2.0\,$eV.  In
addition, we have taken $g=1$.  We have checked that the main features
of all lines of the spectra are independent of the system size.
Hereafter we will take $N=F_{12}=233$ as a representative value.  The
width of the instrumental resolution was $\alpha=0.1\,$eV.  Results at
finite temperature comprise $25$ thermal averages, i.e. averages over
realizations of $V_n$.

\begin{figure}
\centerline{\epsfig{file=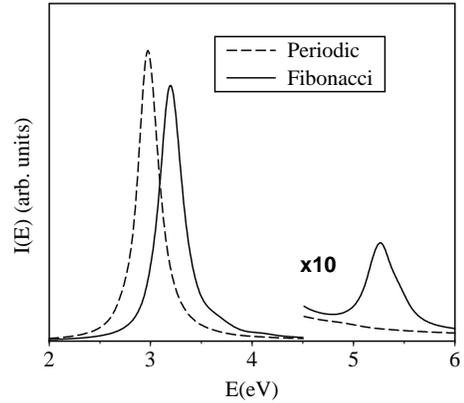,height=6cm}}
\caption{Absorption spectra for a Fibonacci lattice (solid line) and
a pure A lattice (dashed line) at $T=77\,$K.}
\label{fig1}
\end{figure}

In the pure A lattice ($V_n=V_A$) at zero temperature the spectrum
consists of a single Lorentzian line centered at $E^{(A)} = V_A-2J =
3.0\,$eV.  On increasing temperature this line is also Lorentzian-shaped
for weak scattering ($D\ll W$)~\cite{Schreiber82}.  Notice that the weak
scattering case is an excellent approximation in the whole temperature
range of interest.  Figure~\ref{fig1} shows the results for a pure A
lattice at $T=77\,$K.  As soon as B centers are introduced into the lattice
according to the Fibonacci inflation rule, the main absorption line is
shifted towards higher energies.  The value of the energy shift can be
computed using standard perturbation techniques to obtain $\Delta E =
(1-\tau)(V_B-V_A)$~\cite{PLA}.  Therefore, the main absorption line of the
Fibonacci lattice at zero temperature is centered at $E^{(AB)} = (1-\tau)V_B +
\tau V_A - 2J=3.3\,$eV.  Since we are in the weak scattering regime, the
main absorption line is still Lorentzian, as it can be seen in
Fig.~\ref{fig1} at $T=77\,$K.  Besides this main absorption line, a
satellite line can also be observed in the high-energy part of the
spectrum of the Fibonacci lattice, centered at about $5.2\,$eV at zero
temperature (see Fig.~\ref{fig1}).  Using again standard perturbation
techniques \cite{PLA}, it is possible to demonstrate that a series of
satellite peaks appear centered at
\begin{equation}
E_{p}^{(AB)}=E^{(AB)}+4J\sin^{2}\left( \pi \tau^{p} \right),
\label{peaks}
\end{equation}
where $p=2,3,\ldots$ These satellite peaks are the fingerprint of the
Davidov splitting~~\cite{Davidov71} since the Fibonacci lattice can be
regarded as a crystal with a large unit cell.  From (\ref{peaks}) we
obtain $E_{2}^{(AB)}=5.0\,$eV, clse to the satellite observed in 
Fig.~\ref{fig1}.  Similarly $E_{3}^{(AB)} = 4.2\,$eV and so on.  
However, the remainder satellites are hidden by the main absorption
line and cannot be observed in Fig.~\ref{fig1}.

\begin{figure}
\centerline{\epsfig{file=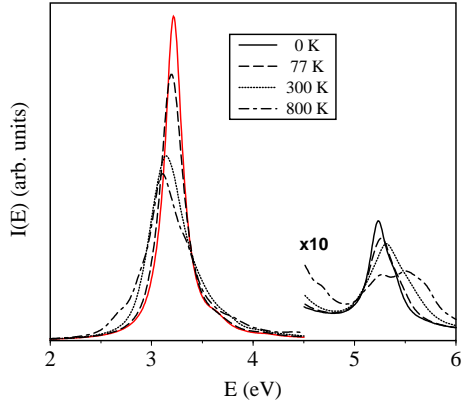,height=6cm}}
\caption{Absorption spectra for a Fibonacci lattice at different
temperatures: $T=0\,$K (solid line), $77\,$K (dashed line),
$300\,$K (dotted line), and $800\,$K (dot-dashed line).}
\label{fig2}
\end{figure}

Figure~\ref{fig2} shows the absorption spectra of the Fibonacci lattice
at different temperatures.  We observe that the main absorption line is
broadened and shifted to lower energies on increasing temperature.  Most
interestingly, the satellite line mentioned above is clearly observed
even at room temperature.  At higher temperature ($T=800\,$K) the
satellite is not well-defined, as can be seen in Fig.~\ref{fig2}.  Some
interesting physical conclusions can be drawn from these results.  First
of all, the quantum coherence required to observe the satellites in the
high-energy region of the spectra is not completely lost even at room
temperature.  The satellite line is caused by the coupling of two
excitonic modes, namely the lowest-lying and that with momentum
$k=\tau^{2}N$, through the topology of the quasiperiodic
lattice~\cite{PLA}.  At zero temperature exciton wave functions extend
over the whole lattice and {\em see} the long-range quasiperiodic order.
On increasing temperature this coherence is partially lost but excitons
still detect the long-range order.  Further temperature increment (say,
above room temperature with our model parameters) yields excitons so
localized that they cannot see the quasiperiodic order.  At this point
it is useful to recall the concept of {\em coherently bound
centers\/}~\cite{Malyshev91,Malyshev95}.  Due to finite temperature
effects, the number of coherently bound centers $N^{*}$ is smaller than the
system size $N$.  This parameter can be determined
from~\cite{Malyshev91,Malyshev95}
\begin{equation}
N^{*}= \left( 3\pi^{2}\,{J\over D} \right)^{2/3} =
       \left( {9\pi^{4}J\over 4gk_BT} \right)^{1/3}.
\label{coherent}
\end{equation}
Since at room temperature the characteristic features of the long-range
order are still present in the optical absorption spectra, whereas at
$T=800\,$K they cannot be detected, we can conclude there exists a
critical value of the number of coherently coupled centers to probe the
long-range order.  We can estimate this critical number from
(\ref{coherent}) to lie between $N^{*}(T=300\,\mathrm{K})=16$ and
$N^{*}(T=800\,\mathrm{K})=12$.

\section{Conclusions}

In summary, we have studied numerically the absorption spectra of
Frenkel excitons on Fibonacci lattices.  Exciton-phonon coupling has
been taken into account because it is a major limiting factor of the
quantum coherence.  This coherence is responsible for satellite lines in
the high-energy region of the optical spectrum.  We have demonstrated
that the characteristic features due to the long-range order of
Fibonacci lattices are observed even at room temperature.  The number of
coherently bound centers needed to probe the long-range order turns out to
be of the order of $N^{*}=15$.  Thus, excitons are indeed a probe of the
underlying quasiperiodicity even at room temperature.  This conclusion
should facilitate future experimental works on linear systems with
long-range order.

\acknowledgments

The authors thank A.\ R.\ Bishop, E.\ Maci\'{a}, V.\ Malyshev and
A.\ S\'{a}nchez for helpful comments and discussions. This work is
supported by CICYT through project MAT95-0325.

\end{document}